\documentstyle[epsfig,12pt]{article}  
\newcommand{\beq}{\begin{equation}}
\newcommand{\eeq}{\end{equation}}
\newcommand{\bea}{\vspace{0.25cm}\begin{eqnarray}}
\newcommand{\eea}{\end{eqnarray}}

\newcommand{\r}{\mbox{{\boldmath
$\rho$}}}

\newcommand{\pb}{{{\bf p}}}

\newcommand{\bb}{{{\bf b}}}

\def\lsim{\mathrel{\rlap{\lower4pt\hbox{\hskip1pt$\sim$}}
    \raise1pt\hbox{$<$}}}         
\def\gsim{\mathrel{\rlap{\lower4pt\hbox{\hskip1pt$\sim$}}
    \raise1pt\hbox{$>$}}}         
\begin{document}
\vspace*{-2cm}
 
\bigskip

\begin{center}

\renewcommand{\thefootnote}{\fnsymbol{footnote}}

  {\Large\bf
Variation of jet quenching from RHIC to LHC and
thermal suppression of QCD coupling constant
\\
\vspace{.7cm}
  }
\renewcommand{\thefootnote}{\arabic{footnote}}
\medskip
{\large
  B.G.~Zakharov
  \bigskip
  \\
  }
{\it
 L.D.~Landau Institute for Theoretical Physics,
        GSP-1, 117940,\\ Kosygina Str. 2, 117334 Moscow, Russia
\vspace{1.7cm}\\}

  {\bf
  Abstract}
\end{center}
{
\baselineskip=9pt
We perform a joint jet tomographic analysis of the data on the 
nuclear modification factor $R_{AA}$ from PHENIX at RHIC and ALICE at LHC.
The computations are performed accounting for radiative and collisional 
parton energy loss with running coupling constant.
Our results show that the observed slow variation of $R_{AA}$ from RHIC to LHC
indicates that the QCD coupling constant is suppressed
in the quark-gluon plasma produced at LHC.
\\
\vspace{.7cm}
}

\noindent{\bf 1.} 
The discovery of strong suppression of high-$p_{T}$ hadrons in
$AA$-collisions  (usually called  the jet quenching)
is one of the main results from the  RHIC program \cite{RHIC_data}.
A similar effect has been observed 
in the ALICE experiment at LHC \cite{ALICE} for $\sqrt{s}=2.76$ TeV. 
The most natural physical reason for this effect is parton energy loss
(radiative and collisional) in the hot quark-gluon 
plasma (QGP) produced in the initial stage of $AA$-collisions.
There has been much progress in the past 15 years in understanding
the radiative energy loss due to induced gluon emission \cite{BDMPS,LCPI,BSZ,W1,GLV1,AMY}.
For RHIC and LHC namely this mechanism dominates the energy loss
\cite{Z_Ecoll}.
Current studies in jet quenching are motivated
by its importance for jet tomography of the dense QCD matter in $AA$-collisions.
Unfortunately, the available approaches to the induced gluon radiation
are limited to one gluon emission.   
While for robust jet tomography, 
one should also take into account multiple gluon emission.
At present multiple gluon radiation is usually accounted for
in the approximation of independent gluon emission \cite{BDMS_RAA}.
However, this approximation has no serious theoretical justification in QCD,
and can lead to considerable systematic errors.
Also, additional uncertainties of the jet tomography come from the unsolved
problem of treatment on an even footing the radiative and collisional (which
is small but not negligible) contributions.

Due to theoretical uncertainties presently
the jet tomography can give only qualitative information about 
density of the QCD matter in $AA$-collisions at a given energy.
However, one may expect that the theoretical uncertainties should not be very 
important for variation of jet quenching with energy. 
For this reason the information extracted from jet tomography based on the
data at very different energies should be more robust.
From this point of view it is very interesting to perform a joint analysis of
the nuclear modification factor $R_{AA}$ 
measured at RHIC and LHC. 
In the present work we perform such an analysis using the 
data on $R_{AA}$ for $Au+Au$ collisions
at $\sqrt{s}=200$ GeV from PHENIX \cite{PHENIX08} and 
for $Pb+Pb$ collisions at $\sqrt{s}=2.76$ TeV from ALICE \cite{ALICE}.
A major purpose of this analysis is to decide whether the variation
of $R_{AA}$ from RHIC to LHC indicates that the QCD coupling constant
becomes smaller in the plasma produced at LHC. The suppression of
the in-medium $\alpha_{s}$ at LHC as compared to that at RHIC energies
would be quite natural since the data on multiplicities 
\cite{N_PHENIX,N_ALICE,N_ALICE2}
indicate that the initial entropy at $\sqrt{s}=2.76$ TeV is bigger by a factor
of $\sim 2.2$ than that at $\sqrt{s}=200$ GeV. This is translated into $\sim 30\%$
growth of the initial temperature at $\sqrt{s}=2.76$ which should lead
to a sizeable thermal suppression of $\alpha_{s}$.
The fact that the PHENIX \cite{PHENIX08} and ALICE \cite {ALICE} 
data on $R_{AA}$ are very similar  supports qualitatively suppression of
$\alpha_{s}$ at LHC.    
It would be interesting to see this from a quantitative analysis.

The method of computation of the nuclear modification factor 
in the present work is similar to that used in \cite{raa08}.
We account for both the radiative and collisional energy losses. They are
calculated with running $\alpha_{s}$ frozen at small momenta. 
We treat the induced gluon radiation
within the light-cone path integral (LCPI) 
approach \cite{LCPI}. The collisional energy loss is viewed as a perturbation
effect.

\vspace{.2cm}
\noindent{\bf 2.}
The nuclear modification factor $R_{AA}$ for a given impact parameter $b$
can be written as
\beq
R_{AA}(b)=\frac{{dN(A+A\rightarrow h+X)}/{d\pb_{T}dy}}
{T_{AA}(b){d\sigma(N+N\rightarrow h+X)}/{d\pb_{T}dy}}\,.
\label{eq:10}
\eeq
Here $\pb_{T}$ is the hadron transverse momentum, $y$ is rapidity (we
consider the central region $y=0$), 
$T_{AA}(b)=\int d\r T_{A}(\r) T_{A}(\r-\bb)$, $T_{A}$ is the nucleus 
profile function. The differential yield for high-$p_{T}$ hadron production in 
$AA$-collision can be written in the form 
\beq
\frac{dN(A+A\rightarrow h+X)}{d\pb_{T} dy}=\int d\r T_{A}(\r)T_{A}(\r-\bb)
\frac{d\sigma_{m}(N+N\rightarrow h+X)}{d\pb_{T} dy}\,,
\label{eq:20}
\eeq
where ${d\sigma_{m}(N+N\rightarrow h+X)}/{d\pb_{T} dy}$ is the medium-modified
cross section for the $N+N\rightarrow h+X$ process.
Similarly to the ordinary pQCD formula, we write it as
\beq
\frac{d\sigma_{m}(N+N\rightarrow h+X)}{d\pb_{T} dy}=
\sum_{i}\int_{0}^{1} \frac{dz}{z^{2}}
D_{h/i}^{m}(z, Q)
\frac{d\sigma(N+N\rightarrow i+X)}{d\pb_{T}^{i} dy}\,.
\label{eq:30}
\eeq
Here $\pb_{T}^{i}=\pb_{T}/z$ is the parton transverse momentum, 
${d\sigma(N+N\rightarrow i+X)}/{d\pb_{T}^{i} dy}$ is the
hard cross section,
$D_{h/i}^{m}$ is the medium-modified fragmentation function (FF)
for transition of a parton $i$ into the observed hadron $h$.
For the parton virtuality scale $Q$ we take the parton transverse
momentum $p^{i}_{T}$.
We assume that hadronization occurs outside of the QGP.
For jets with $E\lsim 100$ GeV the hadronization 
scale, $\mu_h$, is relatively small.
Indeed, 
one can easily show that the $L$-dependence of the parton virtuality
reads $Q^{2}(L)\sim \max{(Q/L,Q_0^{2})}$, where  
$Q_{0}\sim 1-2$ GeV is some minimal nonperturbative scale.
For RHIC and LHC, 
when $\tau_{QGP}\sim R_A$ ($\tau_{QGP}$ is the typical lifetime/size of the
QGP phase, 
$R_A$ is the nucleus radius),
it gives $\mu_{h}\sim Q_{0}$ (for  
$E\lsim 100$ GeV).
Then we can write
\beq
D_{h/i}^{m}(z,Q)\approx\int_{z}^{1} \frac{dz'}{z'}D_{h/j}(z/z',Q_{0})
D_{j/i}^{m}(z',Q_{0},Q)\,,
\label{eq:40}
\eeq
where 
$D_{h/j}(z,Q_{0})$ is the vacuum FF, and
$D_{j/i}^{m}(z',Q_{0},Q)$ is the medium-modified FF
for transition of the initial parton $i$ with virtuality $Q$
to a parton $j$ with virtuality $Q_{0}$.
For partons with $E\lsim 100$ GeV the typical length scale dominating 
the energy loss in the DGLAP stage
is relatively small $\sim 0.3-1$ fm \cite{raa08}.
This length is of the order of the formation time of the QGP
$\tau_{0}\sim 0.5$ fm. On the other hand, the induced radiation stage
occurs at a larger length range $l\sim \tau_{0}\div \tau_{QGP}$. 
For this reason to first approximation
one may ignore the overlap of the DGLAP and induced radiation stages
at all \cite{raa08}. Then we can write
\beq
D_{j/i}^{m}(z,Q_{0},Q)=\int_{z}^{1} \frac{dz'}{z'}D_{j/l}^{ind}(z/z',E_{l})
D_{l/i}^{DGLAP}(z',Q_{0},Q)\,,
\label{eq:60}
\eeq
where $E_{l}=Qz'$,
$D_{j/l}^{ind}$ is the induced radiation FF
(it depends on the parton energy $E$, but not virtuality), and 
$D_{l/i}^{DGLAP}$ is the vacuum DGLAP FF.

We have computed the DGLAP FFs with the help of the PYTHIA event 
generator \cite{PYTHIA}.
Our method of calculation of the in-medium FF via the one gluon 
probability distribution is described in detail in \cite{raa08},
and need not to be repeated here. We just enumerate its basic
aspects:
\begin{itemize}
\item
The multiple gluon emission
is accounted for employing Landau's method as 
in \cite{BDMS_RAA}.
\item
In calculating the $q\to q$ FF the leakage of the probability 
to the unphysical region of $\Delta E> E$ is accounted for by  
renormalizing the FF.
\item
The normalization of the FF for $g\to g$ transition, which does not 
conserve the number of gluons, 
is fixed from the momentum sum rule.
\item
We also take into account the $q\to g$ FF which is usually ignored.
Its normalization
is fixed from the momentum conservation for $q\to q$ and $q\to g$ 
transitions. Thus for quarks our FFs satisfy the flavor and momentum
conservation.   
\end{itemize}

We calculate the hard cross sections  using the LO 
pQCD formula with the CTEQ6 \cite{CTEQ6} parton distribution functions.
To simulate the higher order $K$-factor
we take for the virtuality scale in $\alpha_{s}$ the value 
$cQ$ with $c=0.265$ as in the PYTHIA event generator \cite{PYTHIA}.
We account for the nuclear modification of the parton densities
(which leads to some small deviation of $R_{AA}$ from unity even without
parton energy loss) with the help of the 
EKS98 correction \cite{EKS98}.
For the vacuum FFs we use the KKP parametrization \cite{KKP}.

\vspace{.2cm}
\noindent{\bf 2.}
One gluon induced emission has been computed within the 
LCPI formalism \cite{LCPI}.
The formulas convenient for numerical computations of the induced gluon spectrum
are given in our paper \cite{Z04_RAA}, to which the interested reader is
referred.
We take $m_{q}=300$ and $m_{g}=400$ Mev for the quark and gluon quasiparticle
masses. These values were
obtained in \cite{LH} from the 
analysis of the lattice data within the quasiparticle model for 
the relevant range of the plasma temperature $T\sim (1-3)T_{c}$.
To fix the Debye mass we use the results 
of the lattice calculations for $N_{f}=2$ \cite{Bielefeld_Md} which 
give the ratio $\mu_{D}/T$ slowly decreasing with $T$  
($\mu_{D}/T\approx 3$ at $T\sim 1.5T_{c}$, $\mu_{D}/T\approx 2.4$ at 
$T\sim 4T_{c}$). However, the results for $R_{AA}$ are not very sensitive to
the Debye mass.  

We evaluate the gluon spectrum  with
running $\alpha_s$. The details of incorporating 
the running coupling constant in the LCPI formalism are described
in \cite{Z_Ecoll}. 
We use parametrization of $\alpha_s$ with 
freezing at some value $\alpha_{s}^{fr}$ at 
low momenta. For vacuum a
reasonable choice is $\alpha_{s}^{fr}\approx 0.7$.
This value was previously obtained 
by fitting the low-$x$ proton structure function $F_{2}$ within
the dipole BFKL equation \cite{NZ_HERA}. 
A similar value of 
$\alpha_{s}^{fr}$ follows from the relation 
$
\int_{\mbox{\small 0}}^{\mbox{\small 2 GeV}}\!dQ\frac{\alpha_{s}(Q^{2})}{\pi}
\approx 0.36 \,\,
\mbox{GeV}\,
$
obtained in \cite{DKT} from the analysis of the heavy quark energy 
loss in vacuum.
In vacuum the stopping of the growth of $\alpha_{s}$  
at low $Q$ may be caused by the nonperturbative effects \cite{DKT}. 
In the QGP thermal partons can give an additional suppression of 
$\alpha_{s}$ at low momenta ($Q\sim 2-3T$).
To study the role of the in-medium 
suppression of $\alpha_{s}$ we have performed the numerical 
computations for several smaller values of $\alpha_{s}^{fr}$. 
We are fully aware that this procedure, in which
the in-medium suppression of the coupling is enforced on 
the average, so to speak, via 
modification of one parameter $\alpha_{s}^{fr}$ at any plasma temperature is
very crude. 
We leave more accurate
calculations with a temperature dependent parametrization of $\alpha_{s}$
for future work.

The collisional energy loss is small as compared to the radiative one
but not negligible \cite{Z_Ecoll}.
In the present work we treat the effect of the collisional energy loss 
on the nuclear modification factor as 
a perturbation within the method suggested in \cite{raa08}. It 
consists in renormalization of the initial temperature for
the radiative contribution to the in-medium FFs
using the following equation
$$\Delta E_{rad}(T^{\,'}_{0})=\Delta E_{rad}(T_{0})+\Delta E_{col}(T_{0})\,,$$
where $\Delta E_{rad/col}$ is the radiative/collisional energy loss, $T_{0}$
is the real initial temperature of the QGP, and $T^{\,'}_{0}$ is the 
renormalized temperature. As in \cite{raa08} in calculating
$\Delta E_{rad/col}$ we take for the maximum energy loss half of the initial
parton energy.
The renormalized temperature is not very sensitive to 
the choice of the maximum energy loss in $\Delta E_{rad/col}$.
 
We evaluate $\Delta E_{col}$ within a modified Bjorken method \cite{Bjorken1}
with accurate kinematics of the binary collisions 
(the details can be found in \cite{Z_Ecoll}).
We use the same infrared cutoffs
and parametrization of the coupling constant for the radiative and collisional
energy loss, which is important for minimizing the theoretical 
uncertainties in the fraction  of the collisional contribution.

We also included the effect of the energy gain due to gluon
absorption from plasma by fast partons. It is done with the same prescription as
for collisional energy loss by renormalizing the plasma temperature. 
However, the effect of the energy gain on $R_{AA}$ is practically negligible
\cite{raa08}, and can be safely neglected.

\vspace{.2cm}
\noindent{\bf 4.} 
We describe the QGP in the Bjorken model \cite{Bjorken2}
which gives $T_{0}^{3}\tau_{0}=T^{3}\tau$. We take $\tau_{0}=0.5$ fm. Note
that our numerical results show that $R_{AA}$ is rather insensitive to 
the precise value of $\tau_{0}$. It is physically due to the fact that 
induced gluon emission, dominating parton energy loss, requires typically a finite formation time
which exceeds considerably the value of $\tau_{0}$.
For a given impact parameter the entropy density has been evaluated in the
Glauber model using the Woods-Saxon nucleus density
with parameters as in \cite{ALICE}, and $\sigma_{NN}^{in}=42$ mb for
$\sqrt{s}=200$ GeV and $\sigma_{NN}^{in}=64$ mb for $\sqrt{s}=2.76$ TeV. 
In calculating the entropy distribution in
the impact parameter space we take it in the form 
$dS/dy=C[\alpha dN_{part}/d\eta+(1-\alpha)dN_{coll}/d\eta]$, where the 
$N_{part}$ and $N_{coll}$ terms correspond to the soft and hard mechanisms.
We take $\alpha=0.85$ obtained in the hydrodynamical simulations of $AA$-collisions in \cite{Heinz}. 
The normalization of the entropy has been 
fixed using the entropy/multiplicity ratio
$S/N_{ch}=7.67$ from \cite{BM-entropy}. Note that for $Au+Au$ collisions at
$\sqrt{s}=200$ GeV this prescription gives the total entropy smaller by $\sim$
20\% than that used in \cite{Heinz}.
To simplify numerical computations
for each impact parameter $b$ we calculate the in-medium FFs
for a uniform distribution of the initial temperature $T_{0}$
which was obtained by averaging the realistic entropy distribution computed 
within the Glauber model.
For the 0-5\% centrality bin in $Au+Au$ collisions at $\sqrt{s}=200$ GeV 
we use $dN_{ch}/d\eta=687$
\cite{N_PHENIX},
and in $Pb+Pb$ collisions at $\sqrt{s}=2.76$ TeV $dN_{ch}/d\eta=1601$
\cite{N_ALICE2}.
These multiplicities give $T_{0}\approx 300$ MeV
for the central $Au+Au$ collisions at $\sqrt{s}=200$ GeV
and $T_{0}\approx 400$ MeV
for $Pb+Pb$ collisions at $\sqrt{s}=2.76$ TeV.

The fast parton path length in the QGP, $L$, 
in the medium has been calculated according to the position
of the hard reaction in the impact parameter plane.
For $L>\tau_{QGP}$ we treat the medium as a mixture of plasma and hadron 
phases with relative fractions defined from the decrease of entropy
density as $s\propto 1/\tau$ \cite{Bjorken2}.
To take into account the fact that at times about $1-2$ units of 
$R_{A}$ the transverse expansion
should lead to fast cooling of the hot QCD matter \cite{Bjorken2} we also 
impose the condition $L< L_{max}$. We performed the computations for 
$L_{max}=8$ and 10 fm.  The difference between these two versions is small.

\vspace{.2cm}
\noindent{\bf 5.}
In Fig.~1 the theoretical $R_{AA}$
for $\pi^{0}$ production in the 0-5\% central $Au+Au$ collisions
at $\sqrt{s}=200$ GeV computed with $\alpha_{s}^{fr}=0.7$, 0.6, and 0.5
for chemically equilibrium 
and purely gluonic plasmas is compared to the PHENIX data \cite{PHENIX08}.
The results are presented for purely radiative mechanism
and with inclusion of collisional energy loss and radiative energy
gain. As was said above, the effect of the radiative energy gain on 
$R_{AA}$ is practically negligible.
The growth of $R_{AA}$ for gluons in Fig.~1 is due to the $q\rightarrow g$ 
transition which is usually neglected. 
However, it does not affect strongly the total
$R_{AA}$ since for $\sqrt{s}=200$ GeV 
the gluon contribution to the hard cross section is small at $p_{T}\gsim
15$ GeV.
In Fig.~2 we compare our results for $\alpha_{s}^{fr}=0.7$, 0.5, and 0.4
with the ALICE data 
\cite{ALICE} for charged hadrons in $Pb+Pb$  collisions at $\sqrt{s}=2.76$ TeV. 

As can be seen from Figs.~1,~2, the collisional energy loss suppresses  
$R_{AA}$ only by about 15-25\%.
For the equilibrium plasma the data for $\sqrt{s}=200$ GeV can be 
described with $\alpha_{s}^{fr}\approx 0.6\div 0.7$. 
The data for 
$\sqrt{s}=2.76$ TeV agree better 
with $\alpha_{s}^{fr}\approx 0.4\div 0.5$. 
It provides evidence for the thermal suppression of $\alpha_{s}$
at LHC due to higher temperature of the QGP.
The in-medium suppression of $\alpha_{s}$ 
as the physical reason for a qualitative similarity of $R_{AA}$ at RHIC and LHC
was also recently discussed  in \cite{HG}. However, the effect has not been
investigated quantitatively.

Note that in our previous analysis  \cite{raa08} of the PHENIX data
\cite{PHENIX08} the agreement with experiment was better for
$\alpha_{s}^{fr}\approx 0.5$.
The difference with the present study is due to a somewhat larger total
entropy of the QGP used in \cite{raa08}.
In \cite{raa08} we used the entropy from \cite{Heinz}. 
As was said, the total entropy obtained using $S/N_{ch}$ ratio from
\cite{BM-entropy} is smaller by $\sim 20$\% than that in \cite{Heinz}.
However, this circumstance is not important from the point of view 
the conclusion on suppression of $\alpha_{s}$ at LHC, since we use the
same $S/N_{ch}$ ratio for RHIC and LHC.

\vspace{.2cm}
\noindent {\bf 6}. 
In summary, we have analyzed the 
data on the nuclear modification factor obtained at RHIC in the PHENIX experiment on $Au+Au$ collisions
at $\sqrt{s}=200$ GeV \cite{PHENIX08} and at LHC in the ALICE experiment on
$Pb+Pb$ collisions
at $\sqrt{s}=2.76$ TeV \cite{ALICE}.
The PHENIX data may be described in the scenario with chemically equilibrium
plasma
without (or small) in-medium suppression of $\alpha_{s}$. However, the data on
$R_{AA}$ from ALICE, which are qualitatively similar to that from PHENIX, 
agree better with the results for $\alpha_{s}$ smaller by $\sim 20-30$\% than
that obtained from the PHENIX data.
Thus, a relatively slow variation
of $R_{AA}$ from RHIC to LHC energies indicates that the QCD coupling constant
becomes smaller in the hotter QGP at LHC.

\vspace {.7 cm}
\noindent
{\large\bf Acknowledgements}

\noindent
I am grateful to Jacek Otwinowski for providing me with the 
ALICE data shown in Fig.~2.

\vskip .5 true cm

\newpage
\vspace{-.5cm}
\begin{center}
{\Large \bf Figures}
\end{center}
\begin{figure} [ht]
\begin{center}
\epsfig{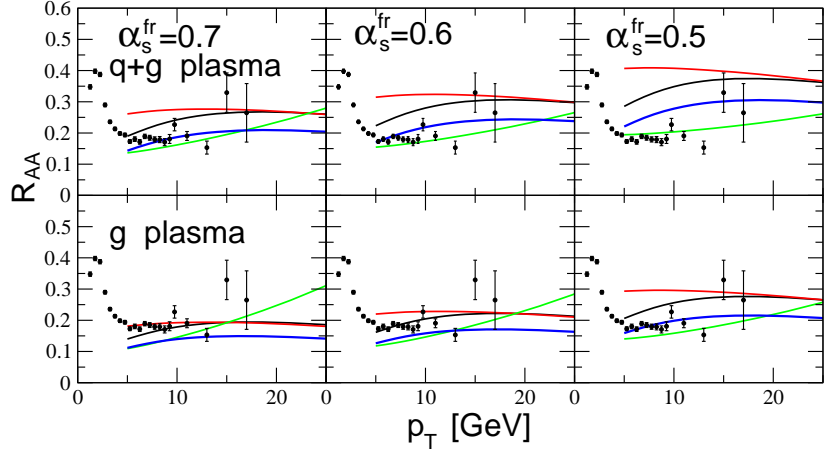}
\end{center}
\caption[.]
{
The factor $R_{AA}$ for $\pi^{0}$ production
in the 0-5\% central $Au+Au$ collisions
at $\sqrt{s}=200$ GeV for $\alpha_{s}^{fr}=0.7$, 0.6, and 0.5.
The upper panels are for the chemically equilibrium
plasma, and the lower ones for purely gluonic plasma.
Black line: the total radiative part (quarks plus gluons);
red line: the radiative quark part;
green line: the radiative gluon part;
blue line: the radiative (quarks and gluons) 
plus collisional, and plus 
energy gain due to gluon absorption.
The theoretical curves obtained for $L_{max}=8$ fm.
The experimental points are the PHENIX data \cite{PHENIX08}.
}
\end{figure}
\begin{figure} [h]
\begin{center}
\epsfig{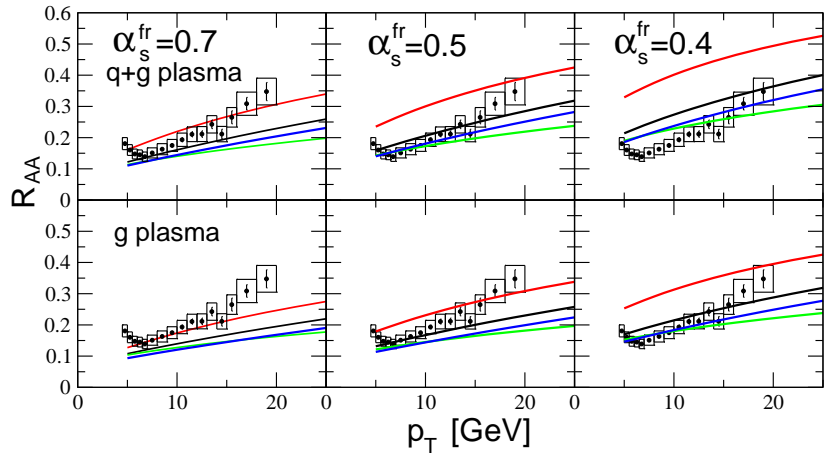}
\end{center}
\caption[.]
{
The same as in Fig.~1 for the charged hadrons in $Pb+Pb$ collisions
at $\sqrt{s}=2.76$ TeV for $\alpha_{s}^{fr}=0.7$, 0.5 and 0.4.
The experimental points are the ALICE data \cite{ALICE},
as in \cite{ALICE} the boxes contain the systematic errors.
}
\end{figure}

\end{document}